\documentclass[aps,preprint]{revtex4}
\usepackage{amssymb}
\usepackage{amsmath}
\usepackage{theorem}
\usepackage{amsfonts}
\usepackage{amsmath}
\usepackage{amssymb}
\usepackage{epsf}
\usepackage{epsfig}
\usepackage{graphicx}
\usepackage{rotating}
\usepackage{boxedminipage}

\begin{document}

\title{Critical dynamics and global persistence in a probabilistic
three-states cellular automaton}
\author{Roberto da Silva (corresponding author)}
\email{rdasilva@inf.ufrgs.br }
\affiliation{Departamento de Inform\'{a}tica Te\'{o}rica, Instituto de Inform\'{a}tica,
Universidade Federal do Rio Grande do Sul. \\
Av. Bento Gon\c{c}alves, 9500, CEP 90570-051 Porto Alegre RS Brazil}
\author{Nelson Alves Junior}
\email{nalves@dfm.ffclrp.usp.br} \affiliation{Departamento de
F\'{\i}sica e Matem\'{a}tica, Faculdade de Filosofia,
Ci\^{e}ncias e Letras, Universidade de S\~{a}o Paulo.\\
Av. Bandeirantes, 3900-CEP 014040-901 Ribeir\~{a}o Preto SP Brazil}

\begin{abstract}
In this work a three-states cellular automaton proposed to describe part of
a biological immune system \cite{tome96} is revisited. We obtain the dynamic
critical exponent $z$ of the model by means of a recent technique that mixes
different initial conditions. Moreover, by using two distinct approaches, we
have also calculated the global persistence exponent $\theta _{g}$, related
to the probability that the order parameter of the model does not change its
sign up to time $t$ [$P(t)\propto t^{-\theta _{g}}$].
\end{abstract}

\maketitle

\setlength{\baselineskip}{0.7cm}

\section{Introduction}

\label{intro}

Cellular automata are statistical-mechanical models which present a complex
behavior, despite their relatively simple dynamic rules. In general these
models are defined in $d$-dimensional lattices with linear dimension $ L $,
in which each site $i$ $(1\leq i\leq L^{d})$ is occupied by a variable $
s_{i}$. The dynamic rules of a typical cellular automaton have two basic
characteristics: Firstly, the update of a variable $s_{i}$ depends only on
its neighborhood (short-range interactions). Secondly, given a configuration
of the system at instant $t$, at instant $t+1$ all variables $s_{i}$ $(1\leq
i\leq L^{d})$ are updated at once. The use of cellular automata in order to
mimic biological immune systems has improved the understanding about the
microscopic mechanisms that lead to the macroscopic behavior of these
systems \cite{bezzi01}. For example, cellular automata are as usefull to
describe the T-helper cells response under parasitic infections \cite%
{brass94,brass94a} as they are to model the course of the Human
Immunodeficiency Virus (HIV) infection \cite{zorzenon01}.

Brass \emph{et al.} divised a simple cellular automaton model of $\mbox{T}_{ %
\mbox{\scriptsize H}}$ cell interactions to mimic the immune system of mice
exposed to parasitic infections \cite{brass94,brass94a}. The model is
defined in a cubic ($d=3$) lattice in which each site is occupied by one of
three different $\mbox{T}_{\mbox{\scriptsize H}}$ cell types: $\mbox{T}_{ %
\mbox{\scriptsize H}}0$, $\mbox{T}_{\mbox{\scriptsize H}}1$ and $\mbox{T}_{ %
\mbox{\scriptsize H}}2$ cells. T-helper cells that have not yet been
presented to the antigen are denoted by $\mbox{T}_{\mbox{\scriptsize H}}0$.
In the model, two distinct routes (equivalent to two populations of antigen
presenting cells) govern the maturing of a $\mbox{T}_{\mbox{\scriptsize H}}0$
cell: Antigen presented by the first (second) route elicit a $\mbox{T}_{ %
\mbox{\scriptsize H}}1$ ($\mbox{T}_{\mbox{\scriptsize
H}}2$) response. In order to take into account the competition between
mature $\mbox{T}_{\mbox{\scriptsize H}}$ cells, the induction $\mbox{T}_{ %
\mbox{\scriptsize H}}0\rightarrow \mbox{T}_{\mbox{\scriptsize H}}1$ ($ %
\mbox{T}_{\mbox{\scriptsize
H}}2$) is forbidden when the neighborhood of the $\mbox{T}_{ %
\mbox{\scriptsize H}}0$ cell has an overall majority of $\mbox{T}_{ %
\mbox{\scriptsize H}}2$ ($\mbox{T}_{\mbox{\scriptsize
H}}1$) cells. At last, a mature cell dies (it is substituted by a $\mbox{T}%
_{ \mbox{\scriptsize H}}0$ cell) if such a cell is not restimulated by the
appropriate antigen within the time interval $N_{T}$. The model can display
a spontaneous symmetry breaking as one varies the antigen density or the
cutoff $N_{T}$, in agreement with experimental results \cite{else93}. A
probabilistic version of the model was proposed by Tom\'{e} and Drugowich de
Fel\'{\i}cio \cite{tome96} by allowing the death of cells $\mbox{T}_{ %
\mbox{\scriptsize H}}1$ and $\mbox{T}_{\mbox{\scriptsize
H}}2$ at each time step with a probability $r$. Thus, in this modified
version of the model, the lifetime $N_{T}$ was substituted by a mean
lifetime related to the probability $r$. Also, the development of $\mbox{T}%
_{ \mbox{\scriptsize H}}0$ cells into either $\mbox{T}_{\mbox{\scriptsize H}%
}1$ or $\mbox{T}_{\mbox{\scriptsize H}}2$ cells occurs with a probability
that depends on the neghborhood of the $\mbox{T}_{\mbox{\scriptsize
H}}0$ cell and on a parameter $p$, related to the antigen density. Such
probabilistic version presents the intrinsic spontaneous symmetry breaking
found in the Brass \emph{et al.} automaton, besides being amenable to an
analytical approach \cite{tome96}. The critical behavior of the
probabilistic model proposed in Ref. \cite{tome96} was studied in a
subsequent work by Ortega \emph{et al.} \cite{ortega98}. By using a
finite-size scaling analysis, from Monte Carlo simulations of square
lattices, Ortega \emph{et al.} determined the ratios of exponents $\frac{
\beta }{\nu }$ and $\frac{\gamma }{\nu }$, as well as the critical point of
the model. Their results suggest that the probabilistic automaton, although
not satisfying detailed balance, belongs to the same universality class of
the two-dimensional kinetic Ising model \cite{ortega98}, thus supporting the
\emph{up-down} conjecture, introduced by Grinstein \emph{et al.} \cite%
{grinstein85}. In a subsequent work, Tom\'{e} and Drugowich \cite{tome98}
obtained the dynamic critical exponent $z$ of the probabilistic automaton in
two dimensions, by performing short-time Monte Carlo simulations by studying
the collapse of the fourth order Binder's cumulant for different lattice
sizes.

In the present work we have revisited the probabilistic automaton of Ref.
\cite{tome96}. We have re-obtained the dynamic critical exponent $z$ of the
automaton by using a recent technique that mixes different initial
conditions \cite{dasilva02}. Such technique is based upon the
short-time critical dynamics introduced by Janssen \emph{et al.}
\cite{janssen89}, who showed that even far from equilibrium the
short-time relaxation of the order parameter follows a universal scale
form
\begin{equation}
M(t)=m_{0}t^{\theta },  \label{m0tatheta}
\end{equation}%
where $M(t)$ is the order parameter at instant $t$ (measured in Monte Carlo
steps per Spin - MCS), $m_{0}=M(0)$ is a small initial $(t=0)$ value of the
order parameter, and $\theta $ is the dynamic critical exponent, related to
the increasing of the order parameter after the quenching of the system
(When dealing with Monte Carlo simulations, the time $t$ is discrete and
measured in Monte Carlo Steps per Spin - MCS). Eq. (\ref{m0tatheta}) demands
working with sharply prepared initial states with a precise value of $m_{0}$%
. After obtaining the critical exponent $\theta $ for a number of $m_{0}$
values, the final value for $\theta $ is obtained from the limit $%
m_{0}\rightarrow 0$.

Starting from an ordered state $(m_0=1)$, the order parameter $M(t)$ decays
in time, at the critical temperature, according to the power law \cite%
{zheng98}
\begin{equation}
\langle M(t)\rangle_{m_0=1}\sim t^{-\beta/\nu z},  \label{bdnz}
\end{equation}
where $\langle (\cdots)\rangle$ is the average of the quantity $(\cdots)$
over different samples with initial order parameter value $m_0=M(0)$, $\beta$
and $\nu$ are the usual static critical exponents, related to the order
parameter and to the correlation length, respectively, and $z$ is the
dynamic critical exponent, defined as $\tau\sim\xi^z$, where $\tau$ and $\xi$
are time and spatial correlation lengths, respectively.

Starting from a disordered configuration with $m_0=0$, the second moment of
the order parameter increases after the power law
\begin{equation}
\langle M^2(t)\rangle_{m_0=0}\sim t^{\omega},  \label{mquad}
\end{equation}
where the exponent $\omega$ is given by
\begin{equation}
\omega=\left (d-\frac{2\beta}{\nu}\right )\frac{1}{z},  \label{omega}
\end{equation}
and $d$ is the dimension of the system.

By combining Eqs. (\ref{bdnz}) and (\ref{mquad}), da Silva \emph{et al.}
\cite{dasilva02} obtained the ratio
\begin{equation}
F_2=\frac{\langle M(t)^2\rangle_{m_0=0}}{\langle M(t)\rangle^{2}_{m_0=1}}
\sim t^{d/z},  \label{f2}
\end{equation}
which corresponds to a function with mixed initial conditions. At this
point, it is important to stress here that, from Eq. (\ref{f2}) above, two
independent runs are necessary in order to calculate the ratio $F_2$: In one
of them $m_0=0$ (numerator), while in the other one $m_0=1$ (denominator).
The ratio $F_2$ has proven to be useful in determining the exponent $z$,
according to recent studies of the 2D Ising model \cite{dasilva02}, $q=3$
and $q=4$ states Potts models \cite{dasilva02}, Ising model with
next-nearest-neighbor interactions \cite{alvesjr03}, Baxter-Wu model \cite%
{arashiro03}, at the tricritical point of the 2D Blume-Capel model \cite%
{dasilva02a}, at the Lifshitz point of the 3D ANNNI model \cite{alvesjr04}
and in other studies concerning models with one absorbent state \cite%
{dasilva04}.

In addition, in this work we have also calculated the global persistence
exponent $\theta_g$ \cite{majumdar96}, related to the probability that the
order parameter of the model, $M(t)$, does not change its sign up to time $t$%
, after a quench of the system to the critical temperature.

The layout of this paper is as follows: In section {\ref{model} we explain
the model. In section \ref{opdcet} we define the order parameter $M(t)$ and
we describe the methodology used in order to obtain the dynamic critical
exponents $\theta_g$ and $z$. In section \ref{results} the results obtained
for the exponents $z$ and $\theta_g$ are shown. Finally, in section \ref%
{conclusion} we present the main concluding remarks. }

\section{The model}

\label{model}

In this work we have studied a two-dimensional probabilistic cellular
automaton in which dynamics is governed by local stochastic rules. At each
site $i$ of the square lattice we have attached a variable $\sigma_i$
assuming the value $0,+1$ or $-1$, depending on whether the site is occupied
by a $\mbox{T}_{\mbox{\scriptsize H}}0$, a $\mbox{T}_{\mbox{\scriptsize H}}1$
or a $\mbox{T}_{\mbox{\scriptsize
H}}2$ cell, respectively. Considering $N$ the total number of sites in the
lattice, we have defined the set $\sigma=(\sigma_1,\sigma_2,\cdots,\sigma_N)$
to represent the microscopic state of the system.

The probability of state $\sigma$ at time $t$, $P_t(\sigma)$, evolves in
time according to the equation
\begin{equation}
P_{t+1}(\sigma^{\prime})=\sum_{\sigma}W(\sigma^{\prime}|\sigma)P_t(\sigma),
\label{evotempprob}
\end{equation}
where the transition probability $W(\sigma^{\prime}|\sigma)$ from state $
\sigma$ to state $\sigma^{\prime}$ obeys the condition
\begin{equation}
\sum_{\sigma^{\prime}}W(\sigma^{\prime}|\sigma)=1.  \label{soma1}
\end{equation}

On the other hand, for a system that evolves at discrete time steps, in
which all the sites are updated at once, as is the case for the cellular
automaton in this work, the transition probability $W(\sigma^{\prime}|
\sigma) $ is written in the form
\begin{equation}
W(\sigma^{\prime}|\sigma)=\prod_{i=1}^{N}\omega_i(\sigma_i^{\prime}|\sigma),
\label{prod1}
\end{equation}
where $\omega_i(\sigma_i^{\prime}|\sigma)$ is the conditional probability
that site $i$ be in the state $\sigma_i^{\prime}$ at time $t+1$, given that
the state of the system is $\sigma$ at instant $t$. Such conditional
probability satisfies the condition
\begin{equation}
\sum_{\sigma_i^{\prime}}\omega_i(\sigma_i^{\prime}|\sigma)=1,  \label{soma2}
\end{equation}
what implies immediately that Eq. (\ref{soma1}) is fulfilled. The cellular
automaton investigated in this work belongs to the class of totalistic
cellular automata \cite{wolfram83}. Thus, the transition probability $
\omega_i(\sigma_i^{\prime}|\sigma)$ depends on $\sigma_i$ and on the sum $
\xi=\sum_{\delta}\sigma_{i+\delta}$, where the sum runs over the
neighborhood of site $i$. More specifically, we are considering a particular
kind of totalistic automaton for which $\omega_i(\sigma_i^{\prime}|\sigma)$
depends only on the sign of the sum $\xi$. By defining
\begin{equation}
s_i=\mbox{sign}(\xi)=\left\{
\begin{array}{lll}
\ \ 1 & \mbox{if} & \xi>0, \\
\ \ 0 & \mbox{if} & \xi=0, \\
-1 & \mbox{if} & \xi<0,%
\end{array}
\right.  \label{sinal}
\end{equation}
we may use the notation $\omega_i(\sigma_i^{\prime}|\sigma_i,s_i)$ in order
to explicit the transition probability dependence on $\sigma_i$ and $s_i$.

The transition probabilities (dynamical rules) are given by
\begin{equation}
\omega_i(+1|\sigma_i,s_i)=p\delta(\sigma_i,0)\{\delta(s_i,+1) +\frac{1}{2}
\delta(s_i,0)\}+(1-r)\delta(\sigma_i,+1),  \label{dinamica1}
\end{equation}
\begin{equation}
\omega_i(-1|\sigma_i,s_i)=p\delta(\sigma_i,0)\{\delta(s_i,-1) +\frac{1}{2}
\delta(s_i,0)\}+(1-r)\delta(\sigma_i,-1),  \label{dinamica2}
\end{equation}
\begin{equation}
\omega_i(0|\sigma_i,s_i)=(1-p)\delta(\sigma_i,0)+r\{\delta(\sigma_i,+1)
+\delta(\sigma_i,-1)\},  \label{dinamica3}
\end{equation}
where, as discussed in section \ref{intro}, $r$ is the death probability of $%
\mbox{T}_{\mbox{\scriptsize H}}1$ and $\mbox{T}_{\mbox{\scriptsize H}}2$
cells, and $p$ is a parameter related to the antigen density \cite{tome96}.

The dynamical rules (\ref{dinamica1}), (\ref{dinamica2}) and (\ref{dinamica3}
) above have ``up-down'', \emph{i.e.},
\begin{equation}
\omega_i(\sigma_i^{\prime}|\sigma_i,s_i)=
\omega_i(-\sigma_i^{\prime}|-\sigma_i,-s_i).  \label{updownsym}
\end{equation}
Therefore, following Grinstein \emph{et al.} \cite{grinstein85}, we expect
that the probabilistic cellular automaton investigated in this work be at
the same universality class of kinetic Ising models.

\section{The order parameter and the dynamic exponent $\protect\theta_g$}

\label{opdcet}

\subsection{Order parameter}

\label{op}

In the short-time Monte Carlo simulations performed in this work the order
parameter $M(t)$, as well as its higher moments $M^{(k)}(t)$, were obtained
from averages over a certain number of samples, $N_s$. By defining the $k$th
momentum of the order parameter in sample number $j$ at instant $t$ as
\begin{equation}
M_{j}^{k}(t)=\frac{1}{L^{2k}}\left(\sum\limits_{i=1}^{L^{2}}\sigma
_{ij}(t)\right) ^{k},  \label{mag_1_amostra}
\end{equation}
the order parameter is written in the form
\begin{equation}
M^{(k)}(t)=\left\langle M_{j}^{k}(t)\right\rangle =\frac{1}{N_{s}L^{2k}}
\sum\limits_{j=1}^{N_{s}}\left( \sum\limits_{i=1}^{L^{2}}\sigma
_{ij}(t)\right) ^{k}.  \label{magmedia}
\end{equation}
As defined in Eqs. (\ref{mag_1_amostra}) and (\ref{magmedia}) above, for $k=1
$ the order parameter is exactly the mean magnetization, $%
M^{(1)}(t)=\left\langle M(t)\right\rangle $.

Eq. (\ref{magmedia}) defines the order parameter and its higher moments for
a set of $N_s$ samples. Thus, if $N_b$ sets of samples are considered at
instant $t$, there are $N_b$ measurements of the magnetization (and its
higher moments), $M_{l}^{(k)}(t)$, where $1\leq l\leq N_b$. By considering
such sets of samples, the final value of the magnetization and its higher
moments are obtained from the average over the $N_b$ sets of samples, \emph{%
i.e.},
\begin{equation}
\overline{M^{(k)}}(t)=(1/N_{b})\sum_{l=1}^{N_{b}}M_{l}^{(k)}(t),
\label{mediasobresets}
\end{equation}
where $M_{l}^{(k)}(t)$ is given by Eq. (\ref{magmedia}) and the
corresponding standard deviation is given by
\begin{equation}
\sigma \left[ M_{l}^{(k)}(t)\right] =\frac{1}{\sqrt{N_{b}(N_{b}-1)^{2}}}
\left( \sum\limits_{l=1}^{N_{b}}\left( M_{l}^{(k)}(t)-\overline{M^{(k)}}
(t)\right) ^{2}\right) ^{1/2}.  \label{desvio_padrao_media}
\end{equation}

\subsection{Dynamic critical exponent $\protect\theta_g$}

\label{dcet}

In this section we define the dynamic critical exponent $\theta_g$ and we
describe two methods used in this work to obtain estimates for $\theta_g$.

In the first method , we have performed short-time Monte Carlo simulations
in order to calculate the global persistence probability $P(t)$, \emph{i.e.}%
, the probability that the magnetization does not change its sign up to time
$t$. On the other hand, the probability $P(t)$ is numerically equal to the
complement of the accumulated distribution $p(t)$, according to which the
magnetization changes its sign for the first time exactly at instant $t$,
\emph{i.e.},
\begin{equation}
P(t)=1-\sum_{t^{\prime }=1}^{t}p(t^{\prime })=1-\sum_{t^{\prime }=1}^{t}
\frac{n(t^{\prime })}{N_{s}},  \label{acumulada}
\end{equation}
where $n(t^{\prime })$ is the number of samples for which the magnetization
changes its sign for the first time at instant $t^{\prime}$ and $N_s$ is the
total number of samples. The exponent $\theta_g$ may be obtained directly
from the power law scale relation \cite{majumdar96}
\begin{equation}
P(t)\sim t^{-\theta _{g}},  \label{powerlawthetag}
\end{equation}
from which we obtain $\ln P(t)=c-\theta _{g}\ln t$, where $c$ is constant
and each run requires a sharply prepared initial state, with a precise small
value of $m_0$, as discussed in Eq. (\ref{m0tatheta}).

In a second way to obtain the exponent $\theta_g$ we have used the fact that
the initial magnetization dependence of $P(t)$ can be cast in the following
finite-size scaling relation \cite{majumdar96},
\begin{equation}
P(t)=t^{-\theta _{g}}\,f(t/L^{z})=L^{-\theta _{g}z}\,{\tilde{f}}(t/L^{z})\,,
\label{leiescalapersistI}
\end{equation}
which renders a different method to obtain the exponent $\theta _{g}$ from
lattice sizes $L_{1}$ and $L_{2}$ \cite{majumdar96}. For this end we define $%
W(t,L)=L^{\theta _{g}z}P(t)$, which turns out to be a function of $t/L^{z}$.
Therefore, if we fix the dynamic exponent $z$, the exponent $\theta_{g}$ can
be obtained by collapsing the time series $W(t_{2},L_{2})=f(t_{2}/L_{2}^{z})$
onto $W(t_{1},L_{1})=f(t_{1}/L_{1}^{z})$ as follows. Under re-scaling, with $%
b=L_{2}/L_{1}$, $(L_{2}>L_{1})$, we obtain
\begin{equation}
W(t_{2},L_{2})=\widetilde{W}(b^{z}t_{1},bL_{1})\ ,
\label{leiescalapersistII}
\end{equation}
and the best estimate for $\theta _{g}$ corresponds to the minimization of
\begin{equation}
\chi ^{2}(\theta _{g})=\sum\limits_{t}\left( \frac{W(t,L)-\widetilde{W}
(b^{z}t,bL)}{\left\vert W(t,L)\right\vert +\left\vert \widetilde{W}
(b^{z}t,bL)\right\vert }\right) ^{2}  \label{desvquadmed}
\end{equation}
by interpolating $\widetilde{W}$ to the time values $b^{z}t$. In order to
obtain the exponent $\theta_g$ using the collapse method described above, it
is not necessary to fix a precise value of the initial magnetization $m_0$
in the short-time simulations, once the scaling relation in Eq. (\ref%
{leiescalapersistI}) does not take into account the initial conditions of
the system. So, we have used initial states in which $\left\langle
m_{0}\right\rangle \sim 0$. On the other hand, the collapse method demands
the dynamic exponent $z$ to be known beforehand. Therefore, in this work we
have used the scaling relation of Eq. (\ref{f2}) in order to obtain
estimates for the exponent $z$. Although both methods described by Eqs. (\ref%
{powerlawthetag}) and (\ref{leiescalapersistI}) were proposed in order to
calculate the exponent $\theta_g$ of the two-dimensional Ising model \cite%
{majumdar96}, such methods were used recently for estimates of $\theta_g$
along the critical line and at the tricritical point of the 2D Blume-Capel
model \cite{dasilva02b}.

\section{Results from short-time Monte Carlo simulations}

\label{results}

In this section we present details about the short-time Monte Carlo
simulations performed for the cellular automaton considered in this work, as
well as the results obtained for both dynamic critical exponents $z$ and $%
\theta_g$.

\subsection{Critical parameters}

\label{critpoints}

\begin{figure}[th]
\centerline{\psfig{file=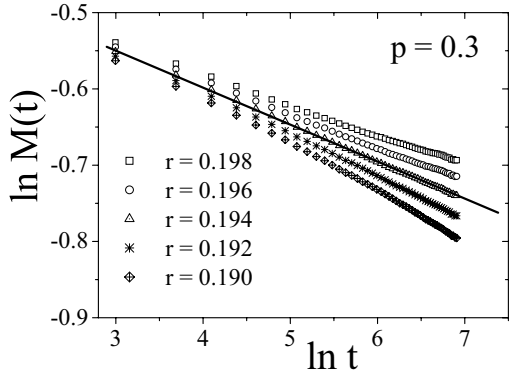,width=8cm}} \vspace*{8pt}
\caption{Log-log plots of the order parameter (magnetization) $M(t)$ versus $%
t$, constructed for $p=0.3$ and different values of $r$. The best linear
fitting yields the best estimate for the critical value of the parameter $r$%
. From the results summarized in Table \protect\ref{table1} (see text), we
have located the critical point $(p=0.3;r=0.194)$.}
\label{fig1}
\end{figure}

Initially we refine the critical parameter $r=0.196$ obtained in Ref. \cite%
{tome96}) for $p=0.3$. From Eq. (\ref{bdnz}) we expected a straight line for
the log-log plot of $M(t)$ against $t$ at the critical point $(p=0.3;
r=0.196)$. However, from log-log plots of Eq. (\ref{bdnz}) obtained for
different values of $r$, we observed that more accurate straight lines are
obtained for $r\neq 0.196$, as depicted in Fig. \ref{fig1} for $%
r=0.190,0.192,0.194,0.196$ and $0.198$. In the short-time simulations
performed in order to obtain the curves shown in Fig \ref{fig1} we have used
square lattices $(d=2)$ with linear dimensions $L=160$, $N_s=10000$ samples,
$N_b=5$ sets of samples and 1000 MCS.
For each value of $r$ used in Fig. \ref{fig1} we calculated the goodness of
fit $Q$ for the same time interval $[t_1,t_2]$ and obtained the values shown
in Table \ref{table1}, from which we obtain the critical value $r=0.194$.
\begin{table}[htb]
\caption{Values of $Q$ for different values of $r$ obtained from linear
fitting of log-log plots of the magnetization $M(t)$ versus $t$.}
\label{table1}\vspace{.3cm}
\par
\begin{center}
\begin{tabular}{|c|c|c|}
\hline\hline
Time interval & $r$ & $Q$ \\ \hline\hline
$[50,300]$ & $0.190$ & $2.31\times 10^{-7}$ \\ \hline
$[50,300]$ & $0.192$ & $0.166$ \\ \hline
$[50,600]$ & $0.194$ & $0.468$ \\ \hline
$[50,300]$ & $0.196$ & $1.04\times 10^{-28}$ \\ \hline
$[50,300]$ & $0.198$ & $1.09\times 10^{-76}$ \\ \hline\hline
\end{tabular}
\end{center}
\end{table}

In order to confirm the critical value obtained in Table \ref{table1}, we
have also calculated the critical value of $r$ for $p=0.3$ by using the
effective exponent, which is given by \cite{grassberger96}
\begin{equation}
\lambda (t)=\dfrac{1}{\Delta t}\log \left( \frac{M(t)}{M(t/\Delta t)}\right).
\label{effexp}
\end{equation}
Such exponent takes into account finite-time corrections (finite-time
scaling) and, in the limit $t\rightarrow \infty$, Eq. (\ref{effexp}) behaves
as \cite{grassberger96}
\begin{equation}
\lambda (t)=c_{1}+c_{2}/t,  \label{asympeffexp}
\end{equation}
where $c_{1}$ and $c_{2}$ are numerical constants and $\Delta t$ is a fixed
time step.

\begin{figure}[th]
\centerline{\psfig{file=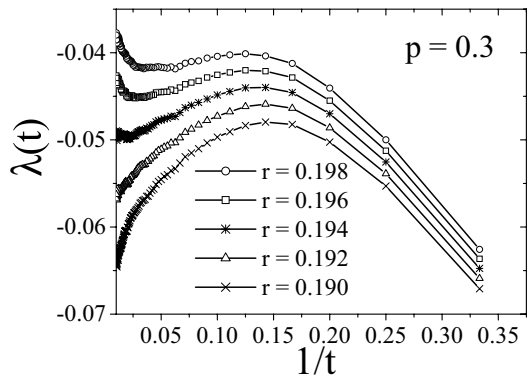,width=8cm}} \vspace*{8pt}
\caption{Effective exponent $\protect\lambda (t)$ obtained for $p=0.3$ and
different values of $r$. From this graph, we have obtained the critical
point $(p=0.3; r=0.194)$ (See text for details).}
\label{fig2}
\end{figure}

Log-log plots of Eq. (\ref{effexp}) are shown in Fig. \ref{fig2}, for $%
\Delta t=10$, $p=0.3$ and $r=0.190,0.192,0.194,0.196$ and $0.198$. From Fig. %
\ref{fig2} it is clear that the asymptotic behavior of $\lambda (t)$ given
by Eq. (\ref{asympeffexp}) is verified for $r=0.194$, thus confirming the
result previously obtained from log-log plots of Eq. (\ref{bdnz}).

Before presenting the estimates obtained for the critical exponents $z$ and $%
\theta_g$ in the next subsections, it is important to emphasize here that
the Monte Carlo simulations were performed only at the critical point $%
(p=0.3; r=0.194)$.

\subsection{Critical dynamic exponent $z$}

\label{calcz}

\begin{figure}[th]
\centerline{\psfig{file=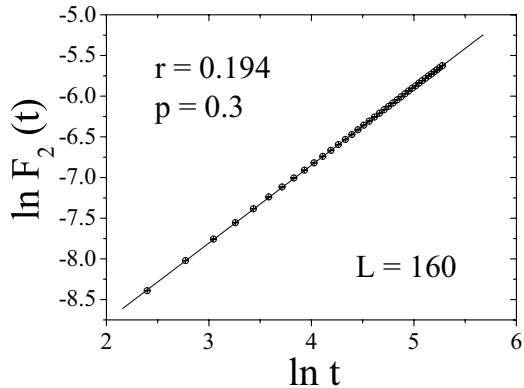,width=8cm}} \vspace*{8pt}
\caption{Typical log-log plot of $F_{2}(t)$ versus $t$ (straight line),
obtained from short-time Monte Carlo simulations at the critical point $%
(p=0.3; r=0.194)$. }
\label{fig3}
\end{figure}

In order to obtain the dynamic critical exponent $z$ we have perfomed
short-time Monte Carlo simulations for square lattices $(d=2)$ with linear
dimensions $L=160$, $N_{s}=20000$ samples, $N_{b}=5$ sets of samples and 200
MCS. From the definition of the ratio $F_2$ given by Eq. (\ref{f2}), we have
constructed the log-log plot of $F_2(t)$ versus $t$ shown in Fig. \ref{fig3}%
, obtained from Monte Carlo simulations performed for one set of samples.
The straight line depicted in Fig. \ref{fig3} is in accordance with the
linear behavior predicted in the scaling relation of Eq. (\ref{f2}).

In Table \ref{table2} we summarize the estimates for $z$ obtained from
different time intervals $[t_1,t_2]$, together with the corresponding values
of $Q$.
\begin{table}[htb]
\caption{Dynamic critical exponent $z$ and the corresponding value of $Q$
obtained for different time intervals.}
\label{table2}\vspace{.3cm}
\par
\begin{center}
\begin{tabular}{|c|c|c|}
\hline\hline
Time interval & $z$ & $Q$ \\ \hline\hline
$[10,120]$ & $2.0853(7)$ & $0.2$ \\ \hline
$[10,160]$ & $2.0856(5)$ & $0.49$ \\ \hline
$[10,200]$ & $2.0869(4)$ & $0.54$ \\ \hline
$[30,200]$ & $2.084(4)$ & $0.98$ \\ \hline
$[100,200]$ & $2.097(8)$ & $0.99$ \\ \hline\hline
\end{tabular}
\end{center}
\end{table}

From Table \ref{table2}, we have obtained the best estimate for the dynamic
critical exponent $z=2.097(8)$, once the goodness of fit $Q=0.99$ is maximum
at the corresponding time interval. However, such value of $z$ is somewhat
smaller than $z=2.17(2)$, which was obtained from the fourth order Binder's
cumulant \cite{tome98}. In order to check the value of $z$ obtained in this
work, in the next section we obtain the global persistence exponent from the
collapse method, which depends on the value of $z$, and directly from Eq. (%
\ref{powerlawthetag}), which does not depend on the value of $z$. As we
shall see in the following, the results for $\theta_g$ obtained from these
two approaches are in very good agreement with each other.

\subsection{Critical dynamic exponent $\protect\theta_g$}

\label{calcthetag}

\begin{figure}[htb]
\centerline{\psfig{file=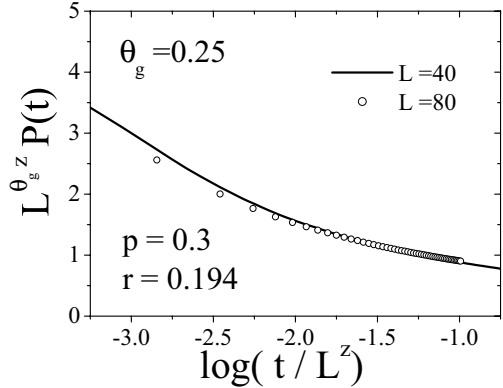,width=8cm}} \vspace*{8pt}
\caption{Collapse obtained from the scaling relation Eq. (\protect\ref%
{leiescalapersistI}), corresponding to square lattices with linear
dimensions $(L_1,L_2=40,80)$ and $\protect\theta_g=0.25$.}
\label{fig4}
\end{figure}

By using Eqs. (\ref{leiescalapersistI}), (\ref{leiescalapersistII}) and (\ref%
{desvquadmed}) given in section \ref{dcet}, we have performed short-time
Monte Carlo simulations in order to apply the collapse method of section \ref%
{dcet} and obtain the dynamic critical exponent $\theta_g$. According with
the description of the method, the dynamic exponent $z$ is to be known
beforehand. Thus, we have used the value $z=2.097$ obtained in the previous
section. Monte Carlo runs were made up to 1000 MCS for $N_{s}=40000$ samples
in square lattices with linear dimensions 20, 40 and 80. The collapse method
was applied for pairs of linear dimensions $(L_1,L_2)=(20,40)$ and $%
(L_1,L_2)=(40,80)$, from which we have obtained $\theta _{g}=0.24(2)$ and $%
\theta _{g}=0.25(2)$, respectively. In Fig. \ref{fig4} we show the collapse
of the curves obtained for $L=40$ and $L=80$ with $\theta_g=0.25$.

\begin{figure}[htb]
\centerline{\psfig{file=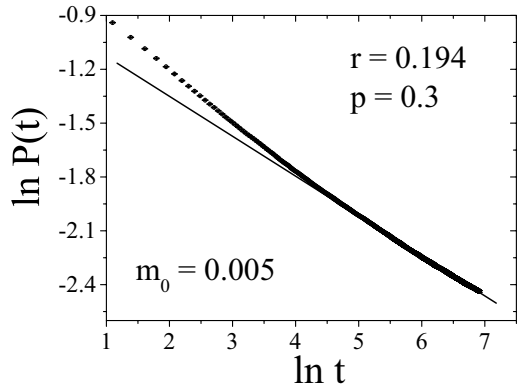,width=8cm}} \vspace*{8pt}
\caption{Typical decay of the probability $P(t)$ (straight line in a log-log
plot), obtained from short-time Monte Carlo simulations of $N_s=40000$
samples with initial state $m_{0}=0.005$.}
\label{fig5}
\end{figure}

Finally, we have obtained the dynamic exponent $\theta_g$ directly from the
power law predicted in Eq. (\ref{powerlawthetag}). To this end, we have
performed short-time Monte Carlo simulations to obtain the quantity $P(t)$,
from which we have constructed log-log curves of $P(t)$ versus $t$, as
depicted in Fig. \ref{fig5}.
From Eq. (\ref{powerlawthetag}), the exponent $\theta_g$ was obtained
directly from the slope of such curves. Monte Carlo simulations were
performed for square lattices with linear dimension $L=80$ and $N_s=40000$
samples, where each sample began from the initial state $m_{0}=0.005$. Error
bars were estimated from the Monte Carlo runs performed for $N_b=5$ sets of
samples. In Table \ref{table3} we present the results obtained for the
exponent $\theta_g$ and the corresponding goodness of fit $Q$ for three
different time intervals.
\begin{table}[htb]
\caption{Dynamic critical exponent $\protect\theta_g$ and the corresponding
value of $Q$ obtained for different time intervals.}
\label{table3}\vspace{.3cm}
\par
\begin{center}
\begin{tabular}{|c|c|c|}
\hline\hline
Time interval & $\theta_g$ & $Q$ \\ \hline\hline
$[90,300]$ & $0.247(4)$ & $0.97$ \\ \hline
$[90,400]$ & $0.242(3)$ & $0.86$ \\ \hline
$[100,500]$ & $0.238(6)$ & $0.95$ \\ \hline\hline
\end{tabular}
\end{center}
\end{table}
From Table \ref{table3} above, the best estimate for the global persistence
exponent is $\theta_g=0.247(4)$, corresponding to $Q=0.97$. This result is
in very good agreement with the estimates $\theta_g=0.24(2)$ and $%
\theta_g=0.25(2)$ obtained from the collapse method.

\section{Conclusions}

\label{conclusion}

In this work we have revisited the probabilistic cellular automata
proposed by Tom\'e and Drugowich \cite{tome96} on the basis of a
previous cellular automaton devised by Brass \emph{et al.}
\cite{brass94}. From short-time Monte Carlo simulations, we have
obtained the dynamic critical exponent $z=2.097(8)$ by using a
recent technique that mixes different initial conditions
\cite{dasilva02}. Such result is slightly smaller than the value
$z=2.17(2)$ obtained from the fourth order Binder's cumulant
\cite{tome98}.

We have also performed short-time Monte Carlo simulations in order to
obtain estimates for the dynamic critical exponent $\theta_g$, the
global persistence exponent, by using two distinct approaches: The
collapse method described by Eqs. (\ref{leiescalapersistI}),
(\ref{leiescalapersistII}) and (\ref{desvquadmed}), and directly from
the power law scaling given by Eq. (\ref{powerlawthetag}). From the
collapse method, by using square lattices with linear dimensions
$(L_1,L_2)=(20,40)$ and $(L_1,L_2)=(40,80)$, we have obtained
$\theta_g=0.24(2)$ and $\theta_g=0.25(2)$, respectively. Directly from
Eq. (\ref{powerlawthetag}) we have obtained $\theta_g=0.247(4)$, in
very good agreement with the estimates yielded from the collapse
method.

\section*{Acknowledgments}
N. Alves Jr. acknowledges financial support from Brazilian
agencies FAPESP and CAPES. R. da Silva thanks GPPD of the
Institute of Informatics of Federal University of Rio Grande do
Sul (UFRGS), for computational resources.


\newpage

\begin{thebibliography}{99}
\bibitem{bezzi01} M.~Bezzi. \newblock Modeling evolution and immune system
by cellular automata. \newblock {\em Riv. Nuovo Cimento}, 24(2):1--50, 2001.

\bibitem{brass94} A.~Brass, A.~J. Bancroft, M.~E. Clamp, R.~K. Grencis, and
K.~J. Else. \newblock Dynamical and critical behavior of a simple discrete
model of the cellular immune system. \newblock {\em Phys. Rev. E},
50(2):1589--1593, aug 1994.

\bibitem{brass94a} A.~Brass, R.~K. Grencis, and K.~J. Else. \newblock A
cellular-automata model for helper t-cell subset polarization in chronic and
acute infection. \newblock {\em J. Theor. Biol.}, 166(2):189--200, jan 1994.

\bibitem{zorzenon01} R.~M.~Z. dos Santos and S.~Coutinho. \newblock Dynamics
of {HIV} infection: A cellular automata approach. \newblock {\em Phys. Rev.
Lett.}, 87(16):168102--1--168102--4, oct 2001.

\bibitem{else93} K.~J. Else, G.~M. Entwistle, and R.~K. Grencis. \newblock %
Correlations between worm burden and markers of {TH}1 and {TH}2 cell subset
induction in an inbred strain of mouse infected with trichuris-muris. %
\newblock {\em Parasite Immunol.}, 15(10):595--600, oct 1993.

\bibitem{tome96} T.~Tom\'{e} and J.~R. Drugowich~de Fel\'{\i}cio. \newblock %
Probabilistic cellular automaton describing a biological immune system. %
\newblock {\em Phys. Rev. E}, 53(4):3976--3981, apr 1996.

\bibitem{ortega98} N.~R.~S. Ortega, C.~F.~S. Pinheiro, T.~Tom\'{e}, and
J.~R. Drugowich~de Fel\'{\i}cio. \newblock Critical behavior of a
probabilistic cellular automaton describing a biological system. \newblock
{\em Phys. Lett. A}, 233:93--98, aug 1997.

\bibitem{grinstein85} G.~Grinstein, C.~Jayaprakash, and H.~Yu. \newblock %
Statistical {M}echanics of {P}robabilistic {C}ellular {A}utomata. \newblock
{\em Phys. Rev. Lett.}, 55(23):2527--2530, dec 1985.

\bibitem{tome98} T.~Tom\'{e} and J.~R. Drugowich~de Fel\'{\i}cio. \newblock %
Short-time dynamics of an irreversible probabilistic cellular automaton. %
\newblock {\em Mod. Phys. Lett. B}, 12(21):873--879, sep 1998.

\bibitem{dasilva02} R.~da~Silva, N.~A. Alves, and J.~R. Drugowich~de Fel%
\'{\i}cio. \newblock Mixed initial conditions to estimate the dynamic
critical exponent in short-time {M}onte {C}arlo simulation. \newblock {\em
Phys. Lett. A}, 298:325--329, apr 2002.

\bibitem{janssen89} H.~K. Janssen, B.~Schaub, and B.~Schmittmann. \newblock %
New universal short-time scaling behavior of critical relaxation processes. %
\newblock {\em Z. Phys. B}, 73(4):539--549, 1989.

\bibitem{zheng98} B.~Zheng. \newblock Monte {C}arlo simulations of
short-time critical dynamics. \newblock {\em Int. J. Mod. Phys. B},
12(14):1419--1484, jun 1998.

\bibitem{alvesjr03} N~Alves~Jr. and J.~R. Drugowich~de Fel\'{\i}cio. %
\newblock Short-time dynamic exponents of an {I}sing model with competing
interactions. \newblock {\em Mod. Phys. Lett. B}, 17(5--6):209--218, mar
2003.

\bibitem{arashiro03} E.~Arashiro and J.~R. Drugowich~de Fel\'{\i}cio. %
\newblock Short-time critical dynamics of the {B}axter-{W}u model. \newblock
{\em Phys. Rev. E}, 67(4):046123, apr 2003.

\bibitem{dasilva02a} R.~da~Silva, N.~A. Alves, and J.~R. Drugowich~de Fel%
\'{\i}cio. \newblock Universality and scaling study of the critical behavior
of the two-dimensional {B}lume-{C}apel model in short-time dynamics. \newblock
{\em Phys. Rev. E}, 66(2):026130, aug 2002.

\bibitem{alvesjr04} N~Alves~Jr. and J.~R. Drugowich~de Fel\'{\i}cio. %
\newblock Short-time critical dynamics at the {L}ifshitz point of the {ANNNI}
model. \newblock, to be published.

\bibitem{dasilva04} R.~da~Silva, J.~R. Drugowich~de Fel\'{\i}cio, and
R.~Dickman. \newblock, cond-mat/0404065.

\bibitem{majumdar96} S.~N. Majumdar, A.~J. Bray, S.~J. Cornell, and C.~Sire. %
\newblock Global {P}ersistence {E}xponent for {N}onequilibrium {C}ritical {D}%
ynamics. \newblock {\em Phys. Rev. Lett.}, 77(18):3704--3707, oct 1996.

\bibitem{wolfram83} S.~Wolfram. \newblock Statistical-{M}echanics of {C}%
ellular {A}utomata. \newblock {\em Rev. Mod. Phys.}, 55(3):601--644, 1983.

\bibitem{dasilva02b} R.~da~Silva, N.~A. Alves, and J.~R. Drugowich~de Fel%
\'{\i}cio. \newblock Global {P}ersistence {E}xponent of the {T}wo-{D}%
imensional  {B}lume-{C}apel {M}odel. \newblock {\em Phys. Rev. E},
67(5):057102, may 2002.

\bibitem{grassberger96} P.~Grassberger and Y.~Zhang. \newblock ``{S}%
elf-organized'' formulation of standard percolation phenomena. \newblock
{\em Physica A}, 224(1--2):169--179, feb 1996.
\end{thebibliography}
\end{document}